\documentclass[]{spie}  

\usepackage[usenames,dvipsnames]{color}
 \usepackage{float}
 \usepackage{graphicx}
\usepackage[normalem]{ulem}
 \usepackage[dvipsnames]{xcolor}
 \usepackage{adjustbox}
\usepackage{amsmath,amsfonts,amssymb}
\usepackage{graphicx}
\usepackage[colorlinks=true, allcolors=blue]{hyperref}
\usepackage[colorinlistoftodos]{todonotes}
\usepackage{subcaption}
\usepackage{multirow}
\usepackage{xcolor}
\usepackage{multirow}
\definecolor{BLUE}{RGB}{0,0,255}
\usepackage{booktabs}

\usepackage{graphicx}
\usepackage{multirow}

\title{Retrieval-Augmented Vision Foundation Models for Robust Leukemia Cell Classification across Multiple Microscopy Datasets}

\author[a,$\dagger$]{Carlos Zamora}
\author[a,$\dagger$]{Hiram Zuniga}
\author[a,$\ast$]{Ulises Orozco-Rosas}
\author[a]{Kenia Picos}
\affil[a]{CETYS Universidad, Ave. CETYS Universidad No. 4, Fracc. El Lago, C.P. 22210, Tijuana, Baja California, Mexico}

\authorinfo{$\dagger$These authors contributed equally to this work.\\
*Corresponding author(s).\\U. Orozco-Rosas: E-mail: ulises.orozco@cetys.mx\\[4pt]
Author's version of a manuscript accepted for presentation at SPIE Optics
+ Photonics 2026. Copyright 2026 SPIE.}

\begin{document} 
\maketitle

\begin{abstract}
Leukemia cell image classification is challenged by real-world domain shifts from acquisition, staining, illumination, and site protocols, causing single-dataset models to generalize poorly in real clinical scenarios. This work presents a robust framework for leukemia classification across multiple heterogeneous datasets using a two-stage pipeline with a pretrained vision foundation model. Stage 1 performs binary classification (leukemia vs. non-leukemia) and is trained using 122,167 single-cell images. Stage 2 is conditionally applied to Stage 1 positives to perform subtype classification into Acute Lymphoblastic Leukemia (ALL) and Acute Myeloid Leukemia (AML), trained using 69,400 single-cell images. Labels are harmonized across five heterogeneous datasets to enable cross-dataset training, and performance is evaluated on a held-out dataset protocol to assess domain-shift generalization. Within this pipeline, three encoders are benchmarked (DinoBloom, pretrained on single-cell images; BiomedCLIP, pretrained on biomedical data; and CLIP as a general-purpose model) under linear probing, Low-Rank Adaptation (LoRA), and a Retrieval-Augmented Classification (RAC) module that retrieves the top-k most similar cell images to provide cytomorphological grounding. The objective is to quantify how much domain-specific pretraining contributes to performance under domain shift, and whether cost-effective adaptation and retrieval can be a viable alternative to expensive domain-specialized pretraining. The held-out protocol additionally serves as a diagnostic tool, revealing when classification performance is attributable to dataset-specific artifacts rather than to cytomorphological features.
\end{abstract}


\section{INTRODUCTION}
\label{sec:intro}
 
Leukemia is the most common cancer among children and a leading cause of pediatric deaths \cite{gbd_childhood_leukemia}. Early detection of leukemia is important for patient survival, but confirmatory diagnostic methods are expensive and invasive, complicating detection \cite{yan_cas2025}. This work presents a robust framework for leukemia classification across multiple datasets of single-cell microscopy images. Such a framework could offer a non-invasive first-line screening on peripheral blood smear microscopy images that can help guide the need for more invasive and expensive confirmatory diagnostic methods.
 
Deep learning has been widely used for leukemia classification using blood smear microscopy images \cite{review_applbiosci2025,cnn_compbiomed2025,vitcnn_springer2024}. The problem with this classification task is the real-world domain shifts arising from differences in acquisition devices, staining, illumination, and site protocols, which can degrade the performance of models trained on a single dataset~\cite{Lopez-Montiel2025}. This work uses cross-dataset training over five heterogeneous datasets spanning different acquisition protocols, staining, and illumination. Model robustness is then assessed under domain shift.
 
This study proposes a two-stage pipeline using a pretrained vision foundation model backbone with two classification heads, reducing the diagnostic task into two subtasks: detecting whether a patient has leukemia and, if so, determining its subtype. Stage 1 performs binary classification (leukemia vs. non-leukemia) and is trained using 122,167 single-cell images. Stage 2 is conditionally applied to Stage 1 positives and performs subtype classification into Acute Lymphoblastic Leukemia (ALL) and Acute Myeloid Leukemia (AML), trained using 69,400 leukemic cell images. This two-stage pipeline allows Stage 1 to be trained on 122,167 images rather than being limited to datasets with ALL and AML labels, potentially improving generalization under domain shift. To quantify this, the study evaluates on a held-out dataset representing an unseen domain, instead of a test split from a domain seen during model training.
 
\begin{figure}[t]
\centering
\includegraphics[width=\textwidth]{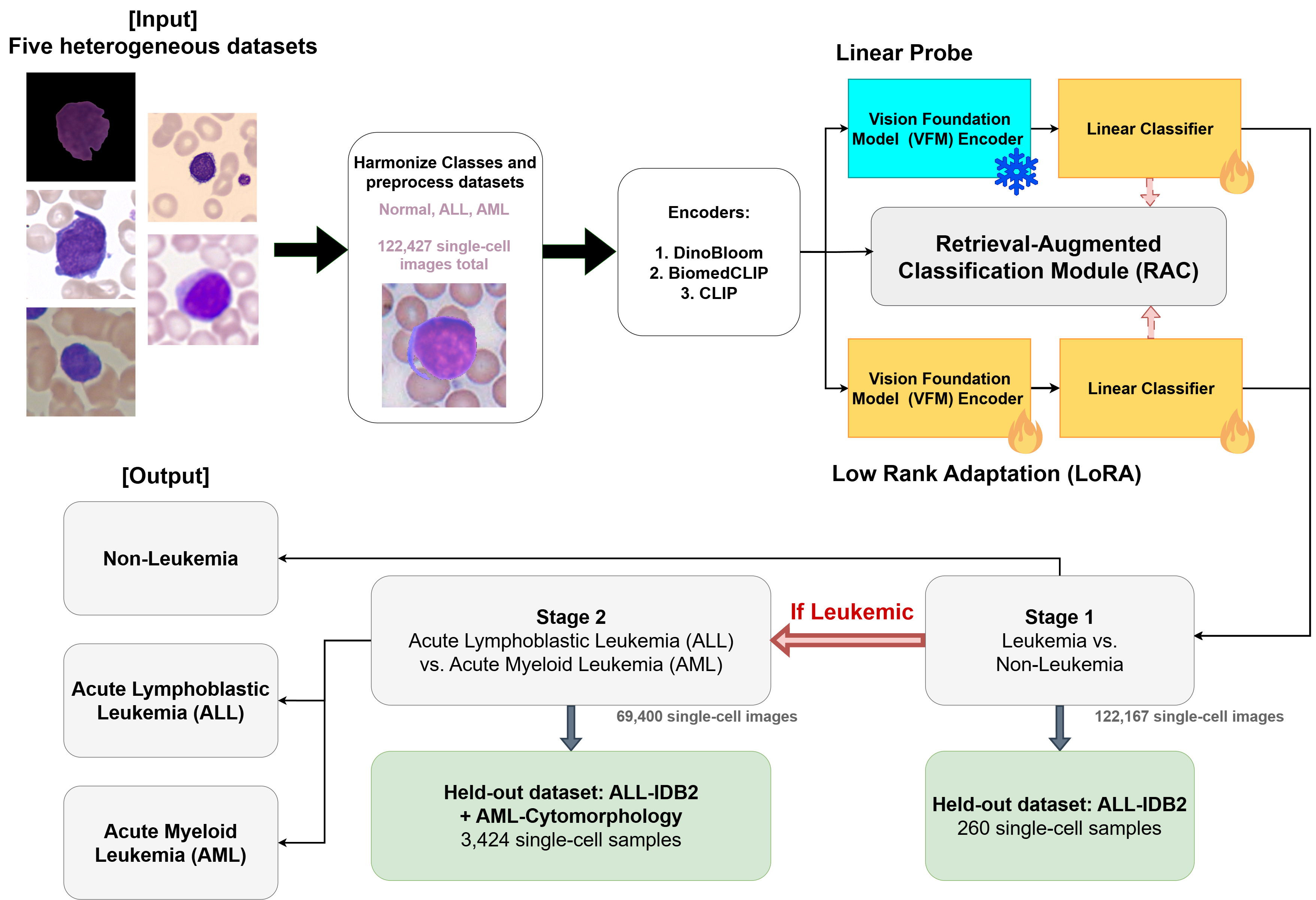}
\caption{Overview of the proposed framework. Five heterogeneous single-cell datasets are harmonized into three classes and encoded by one of three vision foundation models, either frozen with linear probing or adapted with LoRA, with an optional retrieval-augmented classification module. Stage 1 performs binary classification and routes leukemia positives to Stage 2 for subtype classification. Each stage is evaluated on its own held-out dataset. 122,167 and 69,400 are the remaining training data for each stage after excluding the held-out datasets.}
\label{fig:final_pipeline}
\end{figure}

A systematic benchmark is proposed to quantify the impact of specialized and general-purpose Vision Models within the same two-stage pipeline to evaluate how domain-specific pretraining contributes to classification performance. The models used in this work are DinoBloom, which is trained on single-cell images\cite{koch2024dinobloom}, BiomedCLIP as a biomedical model\cite{zhang2025biomedclip}, and CLIP as a general-purpose model\cite{radford2021clip}. Encoder adaptation is assessed through parameter-efficient fine-tuning (PEFT), specifically Low-Rank Adaptation (LoRA). Finally, a Retrieval Augmented Classification (RAC) module that retrieves the top-k most similar labeled cell images in the embedding space is introduced, providing case-based support through cytomorphologically grounded data. An overview of the proposed framework is shown in Fig.~\ref{fig:final_pipeline}.

This work addresses the following research gap: how much does pretraining affect model performance and robustness under domain shift, and how much can fine-tuning and other techniques make up for this difference? Quantifying this impact allows us to evaluate whether cost-effective methods such as PEFT and retrieval augmentation can be a viable alternative to expensive domain-specialized pretraining for leukemia classification.
 
The main contributions of this work are described as follows:
 
\begin{itemize}
\item A two-stage classification pipeline, performing a binary classification at Stage 1 on leukemia vs non-leukemia, and then performing a subtype classification (ALL/AML) on leukemia positives for Stage 2, reducing the broad classification task into two steps.
 
\item A systematic benchmark that quantifies the impact of pretraining specialization on a foundation model for leukemia classification, evaluating models pretrained on single-cell images, biomedical data, and general-purpose data within the same two-stage pipeline. The benchmark further isolates the contribution of adaptation strategies (linear probe and LoRA) and RAC, quantifying how much these techniques compensate for differences in pretraining.
 
\item A held-out dataset evaluation protocol to assess performance under an unseen domain, instead of using a test split from domains seen during training. Cross-dataset training is performed using the remaining heterogeneous datasets (122,167 images for Stage 1 and 69,400 for Stage 2), comprising different acquisition protocols, staining, and labeling paradigms. This evaluation protocol also exposes dependencies on the acquisition source that a within-domain test split cannot reveal.
\end{itemize}
 
The remainder of the paper is structured as follows: Section 2 is related studies; this literature review discusses implementations relevant to our work. Section 3 is the technical background, where foundation concepts specific to this paper are discussed; these are needed to understand our implementation. Section 4 presents the experimental setup, where the methodology and how the results are obtained are presented. Section 5 presents the experimental results and discusses them. Finally, Section 6 presents the Conclusions and Future Work.

\section{Related Studies}
\label{sec:related}
 
Deep learning approaches for leukemia classification have increasingly shifted from Convolutional Neural Networks (CNNs)\cite{lecun1998gradient} to Vision Transformers (ViTs)\cite{dosovitskiy2021vit}, whose self-attention mechanism can capture long-range dependencies and global patterns in cellular features. Prasad and Anbarasi \cite{prasad2024vit} implemented ViTs for ALL detection using the ALL image dataset from the Taleqani Hospital (Tehran, Iran), classifying blood smear images into Benign and ALL morphological types. Similarly, Pandey et al. \cite{praveen2025allvit} used ViTs for the same task using microscopic high-resolution White Blood Cell (WBC) images from blood and bone marrow. Finally, Revathi and Kaliappan \cite{revathi2026leukovit} proposed a hybrid ViT and U-Net framework for ALL classification, trained on single-cell images from the ALL-IDB2 dataset, performing binary classification (ALL or HEM). While these studies demonstrated strong performance on ALL classification, each is limited to only one dataset for training and evaluation, potentially leading to poor generalization under domain shift or cross-dataset evaluation.
 
Various works have acknowledged the importance of cross-dataset training and protocols that may increase model robustness under different acquisition conditions. Preethika and Ananthajothi \cite{preethika2025hybrid} proposed a hybrid model using a ViT-CNN architecture, evaluated using two datasets: ALL-IDB1 and ALL-IDB2, and reporting robustness to image variability. Umer et al. \cite{umer2023imbalanced} discuss robustness to different staining procedures, magnifications, and imaging protocols. They addressed domain shifts using three different datasets to train a robust CNN for WBC classification, addressing missing classes, class imbalance, preprocessing, and so on. Their work uses two datasets we employ, but focuses on white blood cell type classification rather than leukemia detection and subtyping. Wang et al. \cite{wang2026alsnet} proposed ALSNet for morphological cell classification, trained on 180,928 single-cell images from three different imaging platforms to enhance generalizability; they reported good classification results even when using a held-out dataset evaluation, suggesting cross-dataset training from different imaging platforms may increase robustness under domain shift. While this is a key finding for our study, ALSNet focuses on cell-level morphological classification aggregated into case-level subtype predictions, whereas our study trains on a multi-dataset corpus specific to leukemia detection and ALL/AML subtyping. In contrast to these three studies, this work addresses encoder adaptation using LoRA and linear probing, and introduces a RAC module.
 
Outside cross-dataset training, several methods have been explored to improve model generalization in leukemia classification. Patel et al. \cite{patel2024fewshot} investigated the use of vision-language models (VLMs), such as CLIP, BiomedCLIP, and PLIP, for leukemia classification. Their study focused on prompt engineering techniques such as few-shot learning to assess their impact on classification performance using scarce data, showing that biomedical VLMs can achieve strong performance with minimal labeled data. Their work is relevant as it employs general-purpose and biomedical VLMs used in this study, though it does not incorporate encoder adaptation. In another study, the same authors \cite{patel2024evaluating} evaluated the same VLMs for classifying blood cells as normal or malignant, comparing general-purpose VLMs to biomedical-focused VLMs for specialized medical image classification. While this is relevant because it motivates our comparison of encoders, their evaluation does not quantify how adaptation strategies offset pretraining differences.
 
Beyond encoder selection, retrieval-based approaches have also been applied to leukemia. Song et al. \cite{song2026rag} incorporated a knowledge graph-guided RAG from hematology-oncology textbooks for decision support in pediatric acute leukemia; while this implementation is evaluated with text and its purpose is clinical support, rather than performing leukemia classification on images, it serves as a bridge for the RAC modules. Naing et al. \cite{naing2023dml} proposed a deep metric learning system for image retrieval and classification, using a ResNet34 backbone for AML blood cell stages classification, this is a very similar approach to our RAC component because it retrieves images from an embedding space, but differs in that their embedding space is trained from scratch with metric learning objectives, whereas we perform retrieval on pretrained foundation models for cytomorphological grounding data. Another key difference is the use of additional techniques that support this retrieval approach, such as LoRA and linear probe. Their evaluation is limited to a single dataset with an 80/20 split, while we implement cross-dataset training and a held-out dataset evaluation to quantify how RAC contributes to classification performance.
 
While the studies above address individual components in isolation, a few closely related works partially combine them. Dasdelen et al. \cite{dasdelen2026caitomorph} introduced cAItomorph for classifying hematological malignancies from peripheral blood smears, using a DinoBloom encoder with no adaptation techniques or retrieval augmentation. Although trained on a large image corpus comprising eight classes, their external evaluation is restricted to the AML class, leaving the remaining classes untested in external datasets. Ng et al. \cite{ng2026hierarchical} proposed a hierarchical ensemble pipeline for WBC classification under domain shift, using DinoBloom with LoRA adaptation and a retrieval-based inference step conceptually similar to our RAC module. However, their study relies on a single encoder and therefore cannot evaluate how much domain-specific pretraining contributes to performance. Their hold-out test uses the test-split provided by WBCBench, gathered from domains seen during training. Additionally, their focus is WBC classification, not leukemia diagnosis and subtyping.
 
Most closely related to our work, Ben Rabah and Serag \cite{benrabah2026benchmark} proposed a benchmark to quantify the impact of pretraining specialization for computational pathology, evaluating seven general-purpose and three domain-specific foundation models across leukemia, prostate cancer, and breast cancer using linear probing and last-layer fine-tuning adaptation strategies. To assess robustness under domain shift, they trained on a source dataset and evaluated on external datasets for breast and prostate cancer. However, leukemia classification was excluded from this protocol due to inconsistent class definitions across sources, and their four datasets were instead evaluated individually, training and testing within each one. Our work addresses this gap by normalizing labels from five heterogeneous datasets, enabling cross-dataset training and a held-out dataset evaluation protocol. Additionally, LoRA and retrieval augmentation are applied within a two-stage classification pipeline.
 
To the best of the authors' knowledge, no prior work has implemented cross-dataset training through heterogeneous datasets and a held-out dataset test evaluation protocol in the same work for leukemia classification and its subtypes. See Table~\ref{tab:related} for a comparison of related studies against the components integrated in this work. This work presents a framework that unifies all of these components and adds a two-stage pipeline, linear probe, LoRA, retrieval augmentation, and a systematic benchmark. Combined, these components let us answer: how much can cost-effective adaptation techniques improve model performance under domain shift, and assess whether this can be a viable alternative to expensive domain-specialized pretraining of a foundation model, which requires substantial computational resources.

\begin{table}[htbp]
\centering
\caption{Comparison of related studies against the components integrated in this work.
PEFT: parameter-efficient fine-tuning. Enc.: benchmark of encoders with different
pretraining specializations. Retr.: retrieval augmentation. X-train: training across
multiple heterogeneous datasets. Held-out: evaluation on a dataset unseen during training.}
\label{tab:related}
\setlength{\tabcolsep}{3pt}
\begin{tabular}{@{}llccccc@{}}
\toprule
\textbf{Study} & \textbf{Task} & \textbf{Enc.} & \textbf{PEFT} & \textbf{Retr.}
& \textbf{X-train} & \textbf{Held-out} \\
\midrule
Prasad and Anbarasi~\cite{prasad2024vit}      & ALL subtypes        & --  & --  & --  & --  & -- \\
Pandey et al.~\cite{praveen2025allvit}        & ALL subtypes        & --  & --  & --  & --  & -- \\
Revathi and Kaliappan~\cite{revathi2026leukovit} & ALL vs.\ HEM      & --  & --  & --  & --  & -- \\
Preethika and Ananthajothi~\cite{preethika2025hybrid} & Leukemia det. & -- & --  & --  & -- & -- \\
Umer et al.~\cite{umer2023imbalanced}          & WBC typing          & --  & --  & --  & \checkmark & \checkmark \\
Wang et al.~\cite{wang2026alsnet}              & Cell + subtype      & --  & --  & --  & \checkmark & \checkmark \\
Patel et al.~\cite{patel2024fewshot}           & Leukemia det.       & \checkmark & -- & -- & -- & -- \\
Patel et al.~\cite{patel2024evaluating}        & Normal vs.\ malig.  & \checkmark & -- & -- & -- & -- \\
Song et al.~\cite{song2026rag}                 & Clinical QA (text)  & --  & --  & \checkmark & -- & -- \\
Naing et al.~\cite{naing2023dml}               & AML cell stages     & --  & --  & \checkmark & -- & -- \\
Dasdelen et al.~\cite{dasdelen2026caitomorph}  & Malignancy classes  & --  & --  & --  & --  & \checkmark \\
Ng et al.~\cite{ng2026hierarchical}            & WBC typing          & --  & \checkmark & \checkmark & \checkmark & -- \\
Ben Rabah and Serag~\cite{benrabah2026benchmark} & Multi-domain path. & \checkmark & \checkmark & -- & -- & \checkmark$^{\dagger}$ \\
\midrule
\textbf{This work (our proposal)} & \textbf{Leukemia + ALL/AML} & \checkmark & \checkmark & \checkmark & \checkmark & \checkmark \\
\bottomrule
\end{tabular}

\vspace{2pt}
\scriptsize{$^{\dagger}$ Applied to prostate and breast cancer only; hematology was
excluded from the cross-dataset analysis.}
\end{table}

\section{Technical Background}
\label{sec::background}

This section describes the fundamental theory needed to understand the concepts and methods presented in this work.

\subsection{Leukemia and Morphological Diagnosis}
Leukemia is a type of cancer that affects the body’s blood-forming tissues, including the bone marrow. It is characterized by the uncontrolled proliferation of abnormal hematopoietic cells, consequently affecting the production of white blood cells (WBCs), which play a crucial role in protecting against infections. There are several subtypes of leukemia, with the main categories being Acute Lymphoblastic Leukemia (ALL), Acute Myeloid Leukemia (AML), Chronic Lymphocytic Leukemia (CLL), and Chronic Myeloid Leukemia (CML) \cite{li2024leukemiareview}.

Leukemia begins when the DNA of a single hematopoietic precursor cell in a person’s bone marrow mutates, causing its daughter cells to also inherit the same genetic alteration. These abnormal cells accumulate in the bone marrow and eventually enter the bloodstream. The abnormally produced cells are immature and ineffective at supporting the immune system. Furthermore, their uncontrolled proliferation disrupts normal hematopoiesis, which directly affects the body’s ability to produce new healthy blood cells \cite{li2024leukemiareview,passegue2003hematopoiesis}.

Morphological classification of acute leukemias is based on two factors: the cell lineage from which it originates and its maturity stage \cite{ladinescastro2016morphology}. Blood cells develop in the bone marrow and are divided into two main lineages: myeloid (red blood cells, platelets, granulocytes, and monocytes) and lymphoid (B lymphocytes, T lymphocytes, and natural killer cells) \cite{passegue2003hematopoiesis,kondo2010lineage}.

Diagnosing acute leukemias can begin with a non-invasive blood test, where a specialist analyzes a peripheral blood smear sample. The specialist determines whether abnormal cell morphology is present or identifies the presence of leukemic cells \cite{sekar2023morphology}. This process becomes difficult for several reasons, including the morphological similarities between leukemia subtypes, which require analyzing characteristics of the cytoplasm, nucleoli, and nuclear membrane. Additionally, key identifiable cells may appear in multiple leukemia subtypes, and low peripheral blood leukocyte levels can further complicate successful diagnosis \cite{ladinescastro2016morphology}. To assist specialists and potentially improve early screening before invasive diagnostic procedures such as a bone marrow test, computer-assisted diagnostic tools have become a promising tool for leukemia morphological diagnosis.

To prepare and fine-tune an AI-assisted tool, large datasets containing blood cell images are required. Medical image datasets vary in image quality, staining protocols, imaging hardware, and sample sources, introducing varying levels of domain shift, with annotation granularity representing a key characteristic \cite{kumari2025continual}. Some datasets consist of peripheral blood smear images containing WBCs, the primary subjects of study for leukemia diagnosis, and accompanied by surrounding red blood cells (RBCs). These may be annotated at the patient-level or contain additional coordinate-based annotations, requiring an additional cell localization and extraction process before they can be used for morphological classification. In contrast, others provide cropped single-cell images of WBCs, where each cell is individually annotated according to its morphological or diagnostic category \cite{donidalabati2011allidb}.

\subsection{Vision Foundation Models for Medical Image Classification}

Vision Foundation Models (VFMs) are deep learning models that are pretrained on large-scale image datasets to learn visual representations. By being trained on large quantities of diverse, often unlabeled image datasets, these models learn important features and can later be tailored to a specific domain with fine-tuning techniques and adapted to downstream tasks such as image classification~\cite{9490209}. This gained knowledge allows the model to understand the structure of images rather than memorizing patterns from specific tasks, greatly improving its generalization capability and allowing it to reuse these acquired visual representations for various tasks \cite{radford2021clip,oquab2024dinov2}.

Compared with conventional Convolutional Neural Networks (CNNs), which are commonly used in medical imaging for disease classification and grading tasks, including MRI scans, X-rays, and peripheral blood smears \cite{jia2024cnnreview}, VFMs offer increased flexibility and generalization capabilities. Additionally, while supervised CNN-based classifiers are restricted to predefined categories and require task-specific training \cite{lecun1998gradient}, VFMs minimize reliance on task-specific annotated datasets by learning transferable visual representations that can later be adapted to downstream tasks in new domains \cite{radford2021clip,oquab2024dinov2}. This advantage is particularly important in medical imaging tasks, where datasets are scarce, are limited in size, require costly expert annotation, and frequently exhibit significant domain shift due to variations in acquisition methods and medical procedures \cite{kumari2025continual}. Ultimately, VFMs are a promising tool for assisting clinicians in leukemia morphological diagnosis from peripheral blood smear images.

\subsection{Encoder Adaptation Strategies}

Modern AI foundation models, including VFMs, are computationally expensive, and fine-tuning them to a specific problem demands substantial computational resources. Parameter-Efficient Fine-Tuning (PEFT) refers to a collection of methods used to adapt pre-trained deep learning models on new datasets and tasks in a resource- and time-efficient way. PEFT commonly involves freezing most of the model while only training a small subset of parameters, allowing the model to adjust to a new domain while keeping its pretrained knowledge intact. This additionally reduces the risk of catastrophic forgetting, where knowledge that it gained originally is lost and overwritten by newly gained information, all while using significantly less memory and computational power compared to traditional fine-tuning algorithms \cite{han2024peftsurvey}.

Among various PEFT approaches, Low-Rank Adaptation (LoRA) stands out as a resource-efficient way of fine-tuning a model. Unlike traditional techniques, LoRA focuses on freezing the model’s parameters, leaving its pretrained weights intact, and then injecting two new trainable low-rank matrices, B and A, to greatly reduce the number of trainable parameters and decrease the computational power needed for fine-tuning. Instead of using one matrix to represent the full weight update, LoRA approximates it with two smaller ones, using a low-rank decomposition to further reduce the memory needed for this process. This method is based on Eq.~\eqref{eq:lora}.

\begin{equation}
\label{eq:lora}
h = W_0 x + \Delta W x = W_0 x + BAx
\end{equation}

Where $h$ is the resulting output feature representation, $W_0$ is the frozen pretrained weight matrix, $x$ indicates the input feature vector, $\Delta W$ is the trainable low-rank weight update, and $B$ and $A$ are the learnable low-rank matrices where $\Delta W = BA$. This formulation produces the weight updates added to the frozen weights during the forward pass of the model~\cite{han2024peftsurvey,hu2021lora}.

Although VFMs are a breakthrough technology that allows models to learn important visual representations from images, they are expensive to fit to downstream tasks because of their large parameter scale. Medical tasks are no exception, as expertly annotated datasets are limited and typically much smaller than those used during the foundation model pretraining \cite{kumari2025continual}. Consequently, PEFT methods such as LoRA are used to facilitate fine-tuning VFMs to a specific problem, such as leukemia morphological diagnosis. This approach permits efficient adaptation while preserving the powerful visual representations learned during the model’s original large-scale pretraining \cite{zanella2024lowrank}.

\begin{figure}[htbp]
\centering
\includegraphics[width=\linewidth]{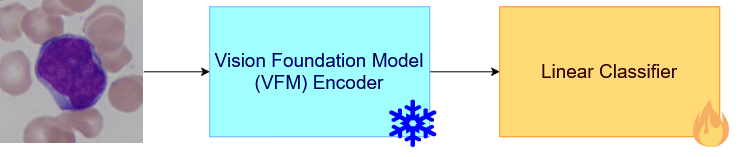}
\caption{Simplified overview of the linear probing process. The pretrained VFM encoder remains frozen while a trainable linear classifier is optimized using the extracted visual representations for the target classification task.}
\label{fig:linear_probe}
\end{figure}

Linear probing for VFMs is a technique commonly used to evaluate the quality of the visual representations learned by the pretrained model. It works by freezing the image encoder and adding a lightweight linear classifier on top. The performance of this classifier shows how well the visual representations transfer to the classification task. High accuracy indicates that these learned representations contain information relevant to the task; otherwise, it suggests that these features are less useful \cite{heim2023domainshift}. A simplified overview of the linear probing process is shown in Fig.~\ref{fig:linear_probe}.

\vspace{0.3cm}

\subsection{Retrieval-Augmented Classification}

Retrieval-augmented methods in artificial intelligence (AI) systems involve implementing an external memory retrieval mechanism for a pretrained model, allowing it to fetch task-specific information and enhance performance \cite{lewis2020rag}. For VFM-based image classifiers, this technique aids the classification process, enabling the model to not only depend on its pretrained visual representations, but also incorporate additional context that helps to accurately classify an image. This is known as Retrieval-Augmented Classification (RAC). The mechanism consists of a parallel retrieval branch that is combined with the image encoder. At inference time, the retrieval branch queries pre-encoded images, which provide additional context and are fused with the base classification to create one final classification \cite{long2022rac}.

\begin{figure}[htbp]
\centering
\includegraphics[width=\linewidth]{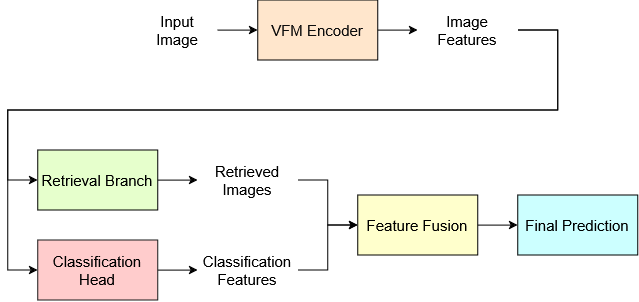}
\caption{Simplified overview of the Retrieval-Augmented Classification (RAC) framework. The input image is processed by a Vision Foundation Model (VFM) encoder, while a parallel retrieval branch retrieves relevant pre-encoded images. The retrieved information is fused with the classification features to produce the final prediction.}
\label{fig:rac_diagram}
\end{figure}

In medical image classification tasks, class imbalance is a commonly observed problem that can create a bias towards the majority (“head”) classes, making it difficult for a model to successfully identify the underrepresented (“tail”) classes \cite{mosquera2024imbalance}. Retrieval-Augmented Classification methods can help in these scenarios, where some disease categories may contain considerably fewer examples; this is also called long-tail classification. By providing additional contextual information, RAC can help the model include information from underrepresented samples during its final classification process, improving its accuracy at classifying “rare” classes \cite{long2022rac}. A simplified overview of the Retrieval-Augmented Classification framework is shown in Fig.~\ref{fig:rac_diagram}.

\vspace{0.3cm}

\section{Experimental Setup}
\label{sec::experimentalsetup}
This section describes the datasets employed in this work, domain shift mitigation, label harmonization, the two-stage classification pipeline, the encoders and adaptation strategies, retrieval-augmented classification, the held-out evaluation protocol and metrics, and the experimental environment.

\subsection{Datasets}
\label{sec:datasets}
Multiple publicly available datasets containing cropped single-cell images of white blood cells (WBCs) were unified into a single dataset, consisting of three main classes: ALL, AML, and normal cells. The purpose of this merge was to improve the framework model's generalization capability by introducing varying image quality, staining protocols, and image acquisition methods, increasing overall sample diversity. After preprocessing, the final dataset contained 122,427 single-cell images.

\subsubsection{C-NMC 2019}
\begin{figure}[H]
    \centering
    \includegraphics[width=\linewidth]{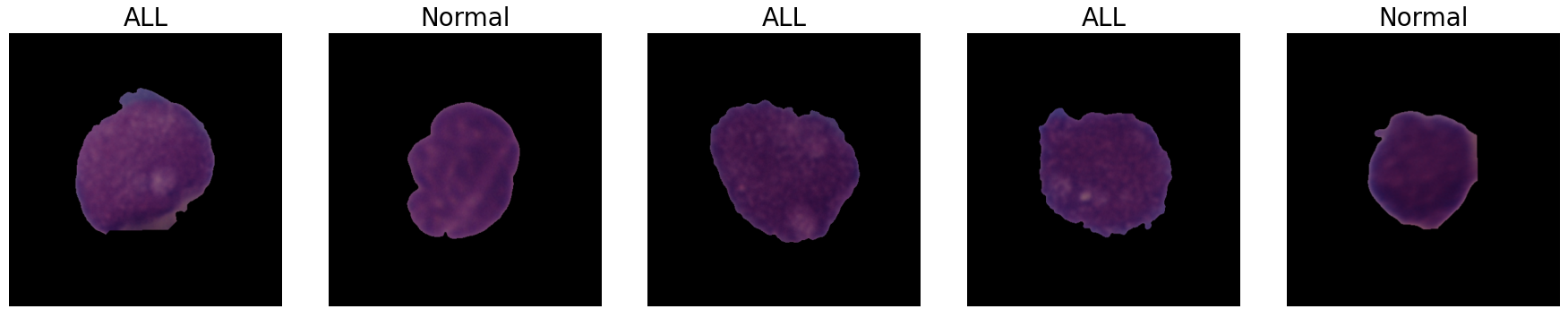}
    \caption{Random samples from the C-NMC 2019 dataset.}
    \label{fig:cnmc_samples}
\end{figure}

The CNMC 2019 dataset was originally released as part of the challenge: Classification of Normal vs Malignant Cells in B-ALL White Blood Cancer Microscopic Images, and contains expert-annotated images of WBCs collected from patients diagnosed with acute lymphoblastic leukemia (ALL) and healthy subjects. The dataset consists of 15,135 cropped single-cell images from 118 subjects, including 69 ALL and 49 healthy patients. Each image contains a segmented stained white blood cell isolated against a black background. Since the official test partition does not provide annotations, only the labeled training and validation subsets were incorporated into the unified dataset used in this work \cite{cnmc2019}. Representative examples of the images contained in the C-NMC 2019 dataset are shown in Fig.~\ref{fig:cnmc_samples}.

\subsubsection{AML-Cytomorphology}
\begin{figure}[H]
    \centering
    \includegraphics[width=\linewidth]{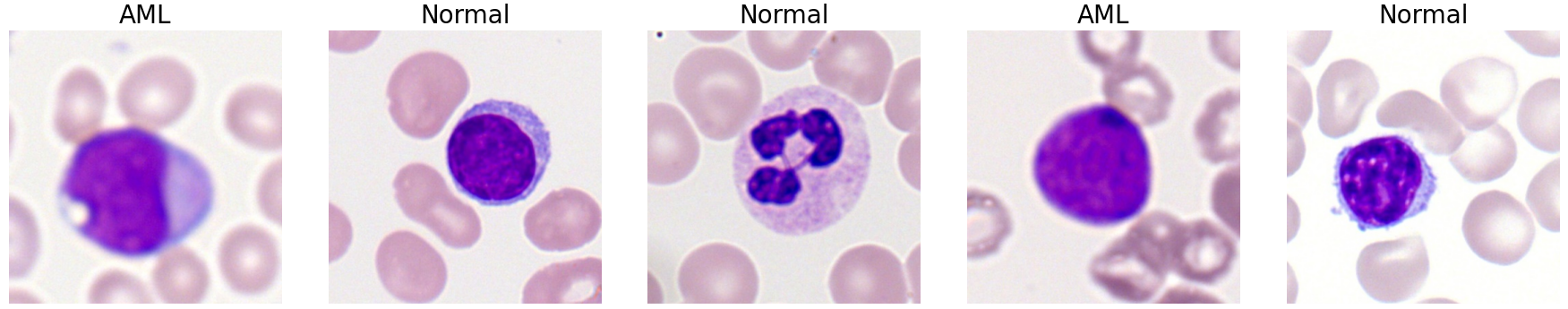}
    \caption{Random samples from the AML-Cytomorphology dataset.}
    \label{fig:aml_cytomorphology_samples}
\end{figure}

The Munich AML-Cytomorphology dataset consists of 18,365 expert-annotated microscopic images of individual white blood cells collected from 200 subjects, including 100 patients diagnosed with acute myeloid leukemia (AML) and 100 healthy patients. The dataset's images are categorized according to cell morphology rather than explicit disease labels. The AML and normal cell labels for this work were derived from the cell morphology annotations. Additionally, ambiguous cell types were excluded, resulting in a curated subset of leukemic and normal cells that was incorporated into the unified dataset \cite{amlcytomorphology}. Representative examples of the images contained in the AML-Cytomorphology dataset are shown in Fig.~\ref{fig:aml_cytomorphology_samples}.

\subsubsection{Peripheral Blood Cell Dataset}
\begin{figure}[H]
    \centering
    \includegraphics[width=\linewidth]{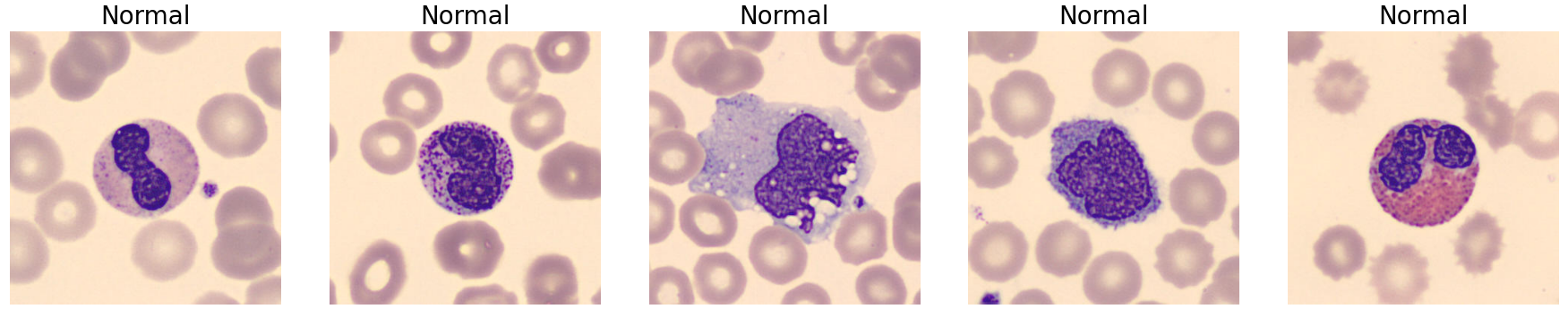}
    \caption{Random samples from the Peripheral Blood Cell Dataset.}
    \label{fig:pbc_samples}
\end{figure}

This dataset contains 17,092 expert-annotated microscopic images of individual normal blood cells acquired from peripheral blood smears. Each sample consists of a cropped single-cell image belonging to one of several normal hematopoietic cell types, including neutrophils, eosinophils, basophils, lymphocytes, monocytes, immature granulocytes, erythroblasts, and platelets. Since the dataset contains only non-leukemic cells, it served as an additional source of normal blood cell images for the unified dataset \cite{acevedo2020wbc}. Representative examples of the normal blood cell images from this dataset are shown in Fig.~\ref{fig:pbc_samples}.

\subsubsection{AML-Cytomorphology MLL Helmholtz}
\begin{figure}[H]
    \centering
    \includegraphics[width=\linewidth]{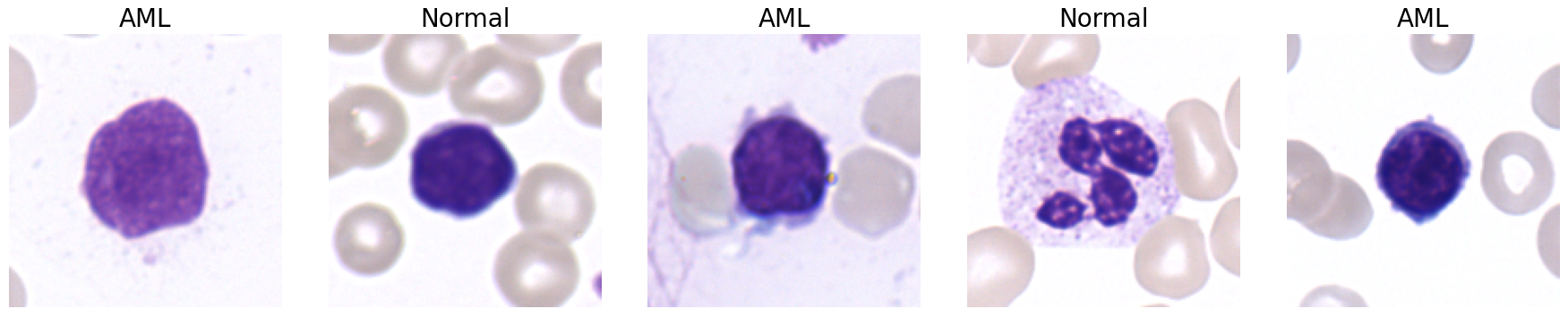}
    \caption{Random samples from the AML-Cytomorphology MLL Helmholtz dataset.}
    \label{fig:aml_mll_samples}
\end{figure}

The AML-Cytomorphology MLL Helmholtz dataset is a large collection of expert-annotated microscopic images of individual white blood cells obtained from peripheral blood smears, acquired from the Munich Leukemia Laboratory (MLL). The dataset consists of 81,214 cropped single-cell images from 189 subjects, including 129  from patients diagnosed with AML and 60 from healthy patients. It includes four genetically defined AML subtypes according to the WHO 2022 classification (PML::RARA, NPM1, CBFB::MYH11, and RUNX1::RUNX1T1) \cite{amlcytomorphologymll}. Representative examples of the images contained in the AML-Cytomorphology MLL Helmholtz dataset are shown in Fig.~\ref{fig:aml_mll_samples}.

\subsubsection{ALL-IDB2}
\begin{figure}[H]
    \centering
    \includegraphics[width=\linewidth]{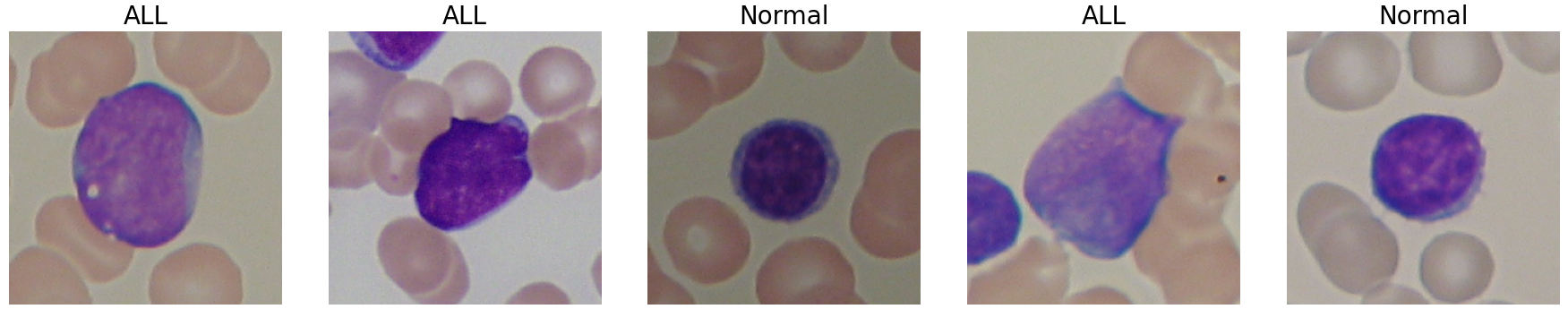}
    \caption{Random samples from the ALL-IDB2 dataset.}
    \label{fig:allidb2_samples}
\end{figure}

The publicly available ALL-IDB2 dataset is designed to evaluate acute lymphoblastic leukemia (ALL) classification algorithms using microscopic blood smear images. The dataset contains 260 expert-annotated cropped single-cell images, equally divided between blast cells from ALL patients and normal white blood cells from healthy subjects. Each image contains a segmented individual cell extracted from peripheral blood smear microscopy. Although relatively small, it was incorporated into the unified dataset to increase the diversity of ALL cell samples \cite{donidalabati2011allidb}. Representative examples of the images contained in the ALL-IDB2 dataset are shown in Fig.~\ref{fig:allidb2_samples}.

\begin{table}[ht]
\centering
\caption{Summary of the datasets used to create the unified dataset. The Original
column reports the size of each dataset as published. The ALL, AML, and Normal
columns report the images retained after applying the inclusion criterion of
Section~\ref{sub:labels}, and therefore sum to the Usable column.}
\label{tab:datasets}
\begin{tabular}{llrrrrrr}
\toprule
& & & & \multicolumn{4}{c}{After filtering} \\
\cmidrule(l){5-8}
Dataset & ID & Subjects & Original & ALL & AML & Normal & Usable \\
\midrule
C-NMC 2019                       & D1 & 118 & 15,135 & 8,491 & --     & 4,037  & 12,528 \\
AML-Cytomorphology               & D2 & 200 & 18,365 & --    & 3,294  & 14,833 & 18,127 \\
Peripheral Blood Cell            & D3 & --  & 17,092 & --    & --     & 10,298 & 10,298 \\
AML-Cytomorphology MLL Helmholtz & D4 & 189 & 81,214 & --    & 60,909 & 20,305 & 81,214 \\
ALL-IDB2                         & D5 & --  & 260    & 130   & --     & 130    & 260 \\
\midrule
\textbf{Total} & & \textbf{507} & \textbf{132,066} & \textbf{8,621} & \textbf{64,203} & \textbf{49,603} & \textbf{122,427} \\
\bottomrule
\end{tabular}

\vspace{2pt}

\end{table}

Table~\ref{tab:datasets} summarizes the original size of each dataset together with the number of images retained after preprocessing and filtering.

\subsection{Domain Shift Mitigation}
One of the main challenges we encountered when creating the unified dataset was the high level of visual inconsistency present in the single-cell images from the C-NMC 2019 dataset. While the other images were similar, consisting of a purple-stained WBC with a light background and often accompanied by red blood cells in the background, those from the C-NMC 2019 dataset were isolated and placed in front of a black background, as seen in Fig.~\ref{fig:cnmc_samples}.

\begin{figure}[H]
    \centering
    \includegraphics[width=\linewidth]{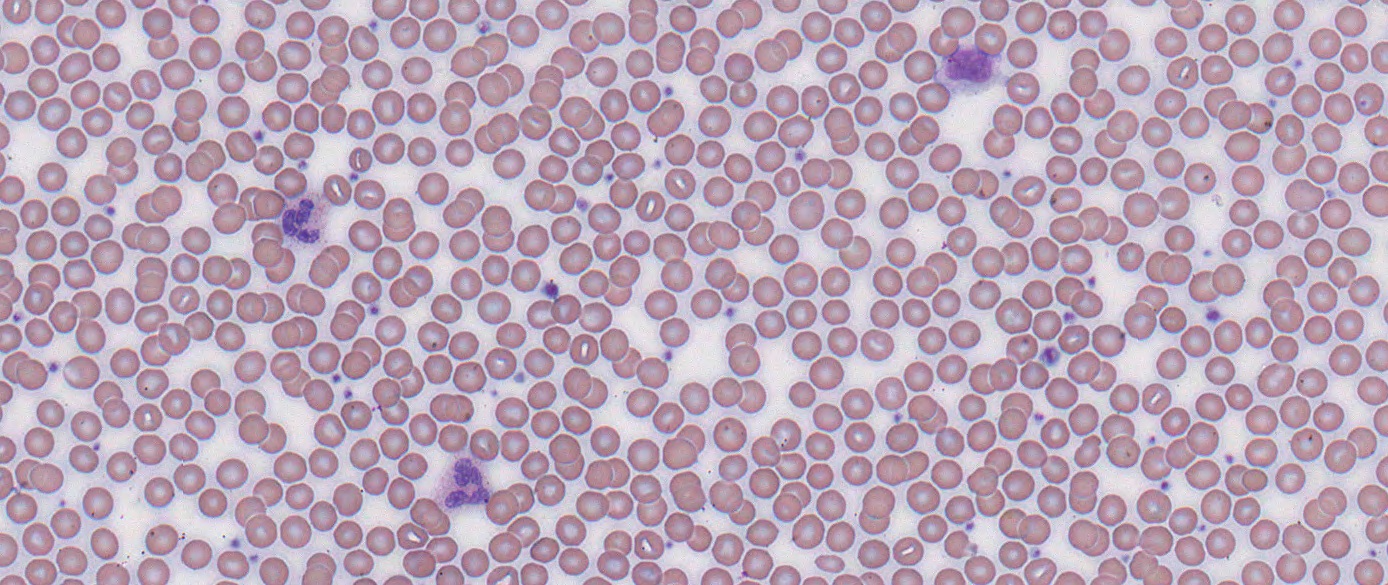}
    \caption{A healthy peripheral blood smear background is used to replace the black backgrounds from C-NMC 2019 images during domain shift reduction.}
    \label{fig:pbs_background}
\end{figure}

To mitigate this large domain shift and reduce the risk of the model learning to differentiate the ALL class by memorizing image elements unrelated to cell morphology, each image was processed. The black background was removed and replaced with a randomly selected area from a healthy peripheral blood smear sample, shown in Fig.~\ref{fig:pbs_background}, containing mainly RBCs \cite{lam2021normal}. To increase generalization, the background was randomly modified through rotations and scaling. Additionally, Reinhard color normalization was applied to improve color consistency between datasets.

\begin{figure}[H]
    \centering
    \includegraphics[width=\linewidth]{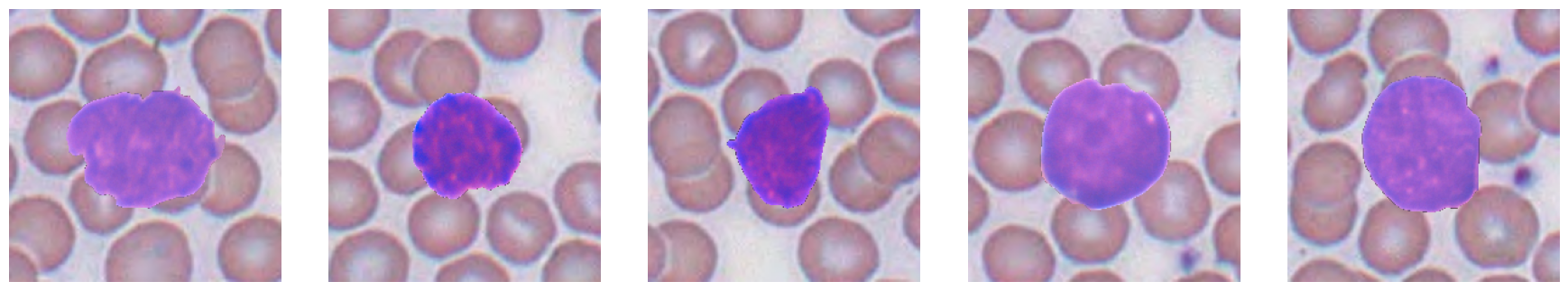}
    \caption{Examples of curated CNMC samples after applying the proposed domain shift reduction technique.}
    \label{fig:curated_cnmc_samples}
\end{figure}

This preprocessing strategy, as seen in Fig.~\ref{fig:curated_cnmc_samples}, helped produce more visually consistent images. While not perfect, the resulting images are similar enough to reduce the likelihood that the model will learn irrelevant image characteristics during training.

\subsection{Label Harmonization}
\label{sub:labels}
Across the five heterogeneous datasets used for training and evaluation, annotation schemes differ substantially. Some provide binary image-level labels (e.g., ALL vs. non-ALL), others provide morphological cell-type labels (e.g., basophil, eosinophil), and others provide AML-type labels (e.g., monoblasts, myeloblasts). Additionally, some datasets are annotated at the patient level rather than the single-cell level (Section~\ref{sec:datasets}). 

To enable cross-dataset training and held-out dataset evaluation, this work harmonizes these heterogeneous labels into three classes: normal, ALL, and AML. All genetic AML subtypes were collapsed into a single AML class. A uniform inclusion criterion was applied: only non-ambiguous nucleated leukocytes were used, since these are the cells relevant to leukemia diagnosis and the only cell type consistently present across all five datasets. Consequently, this study excluded non-leukocyte samples (platelets and erythroblasts), as they are not leukocytes, and smudge cells (labeled as KSC in D2), as they are ruptured cells that cannot be reliably identified. Immature granulocytes (promyelocytes, myelocytes, and metamyelocytes) were excluded because they are labeled inconsistently across datasets. Some sources label them as normal cells and others as leukemic. For the myeloid leukemic class, positive samples were restricted to blasts (myeloblasts and monoblasts). These rules and criteria were applied uniformly across all five datasets to ensure a consistent preprocessing pipeline before training and evaluation.

For the patient-level dataset (D4), cell-level labels are not available because each single-cell crop inherits the diagnostic label of its patient. Since a blood smear from a patient diagnosed with AML still contains a proportion of morphologically normal cells, some individually healthy cells are labeled as AML. This may introduce instance-level label noise. However, it arises from the dataset design rather than from this study’s pipeline. Finally, all images were resized to 224$\times$224 pixels.

\subsection{Two-Stage Classification Pipeline}
\begin{figure}[H]
\centering
\includegraphics[width=\textwidth]{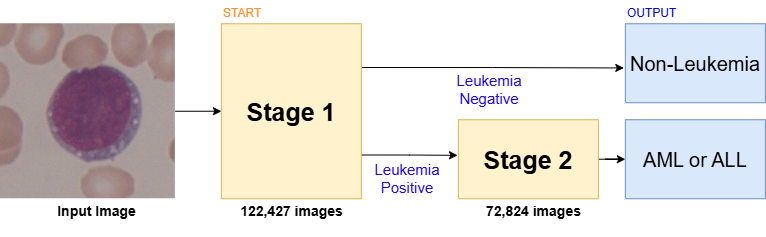}
\caption{Overview of the two-stage classification pipeline. Stage 1 performs binary classification over the full harmonized corpus. Samples predicted as leukemic are passed to Stage 2 for subtype classification. Additionally, 122,427 and 72,824 correspond to the full harmonized corpus per stage, before excluding the held-out datasets.}
\label{fig:pipeline}
\end{figure}

In this work, leukemia classification is decomposed into two sequential stages, each implemented as an independent linear classifier on top of a vision encoder: DinoBloom, BiomedCLIP, or CLIP. Stage 1 performs a binary classification (leukemia vs. non-leukemia) over the full harmonized corpus of 122,427 single-cell images, collapsing the AML and ALL classes into a single leukemia class. Stage 2 performs subtype classification (ALL vs. AML), applied conditionally to the samples predicted as leukemic by Stage 1, over the 72,824 leukemic single-cell images that remain after removing samples labeled as normal. The flow of this pipeline can be seen in Figure \ref{fig:pipeline}.

Decoupling the task this way allows Stage 1 to leverage every labeled image, including datasets that provide only normal single-cell images, rather than being restricted to sources with subtype annotations, augmenting the corpus from 72,824 images to 122,427 images. The two stages are trained and evaluated as separate classifiers under their own held-out datasets (Section~\ref{sec:heldout}), avoiding contamination or data leakage between stages.

\subsection{Encoders and Adaptation Strategies}
Three vision encoders from different pretraining specializations were trained and compared within the same pipeline. DinoBloom\cite{koch2024dinobloom} is a DINOv2-based model \cite{oquab2024dinov2} developed by MarrLab and pretrained on 13 publicly available datasets of single-cell images from peripheral blood and bone marrow, serving as the specialized model. BiomedCLIP\cite{zhang2025biomedclip} is pretrained on biomedical image and text pairs extracted from PubMed Central articles and represents the semi-specialized model, since it is pretrained on a biomedical domain but not as specific as single-cell images. Lastly, CLIP\cite{radford2021clip} is the general-purpose baseline, pretrained on natural image and text pairs. The versions used for these three vision encoders were: DinoBloom-B, which produces 14$\times$14 image patches, 768-dimensional embeddings, and has 85.7 million parameters. BiomedCLIP produces 16$\times$16 image patches, 512-dimensional embeddings, and has 86.2 million parameters, and CLIP produces 32$\times$32 image patches, 512-dimensional embeddings, and has 87.8 million parameters; this information is presented in Table \ref{tab:encoders}. The purpose of choosing these base models was to isolate the pretraining domain as the main source of variation between them, since the three encoders are consistent in parameter quantity; no result can be attributed to encoder size. For the two vision-language models (BiomedCLIP and CLIP), only the vision tower is used, since the pipeline requires no text input. All embeddings are L2-normalized before classification. 

\begin{table}[H]
\centering
\caption{Vision encoders evaluated in this work. All three use the Base Vision Transformer configuration and differ by less than 2.5\% in parameter count, which makes the pretraining domain the main source of variation. Token count corresponds to a $224 \times 224$ input.}
\label{tab:encoders}
\begin{tabular}{lllccc}
\toprule
\textbf{Model} & \textbf{Pretraining domain} & \textbf{Architecture} & \textbf{Parameters} & \textbf{Tokens} & \textbf{Embedding dim.} \\
\midrule
DinoBloom-B & Single-cell hematology  & ViT-B/14 & 85.7M & 256 & 768 \\
BiomedCLIP  & Biomedical image--text  & ViT-B/16 & 86.2M & 196 & 512 \\
CLIP        & Natural image--text     & ViT-B/32 & 87.8M & 49  & 512 \\
\bottomrule
\end{tabular}
\end{table}

Two adaptation strategies are evaluated using these three encoders. Linear probing involves freezing the encoder and training only a logistic regression classifier on the precomputed embeddings, using class weighting to compensate for class imbalance. This configuration isolates the representational quality of the pretrained encoder, since no parameter of the backbone is modified, only the head classifier. No augmentation was applied for linear probing, since embeddings were extracted once with deterministic preprocessing, and the encoder never processes the images again during training.

In Low-Rank Adaptation, low-rank update matrices are injected into the encoder's attention projection layers while the pretrained weights remain frozen. The LoRA parameters and a linear classification head are trained jointly. LoRA was configured with the following hyperparameters: rank $r=8$, scaling factor 16, and dropout 0.05. This results in 296,450 trainable parameters for DinoBloom, 295,938 for BiomedCLIP, and 443,394 for CLIP, a small fraction of the approximately 86 million parameters of each encoder. Training used the AdamW optimizer with a learning rate of $1\times10^{-4}$, a batch size of 128, and class-weighted cross-entropy, for a fixed budget of three epochs with mixed-precision computation. Light augmentation consisting of horizontal and vertical flips and mild color jitter was applied during LoRA training. The same LoRA hyperparameters and training budget were applied uniformly across the three encoders and both stages to ensure comparability. The held-out datasets were never used during training or hyperparameter selection.

\subsection{Retrieval-Augmented Classification}
The RAC module works on the embedding space produced by the frozen or LoRA-adapted encoder. Given a query image, its embedding is compared by cosine similarity against an embedding bank comprising the L2-normalized training embeddings produced by the encoder. The top-$k$ most similar samples make a similarity-weighted vote over classes, making a retrieval distribution $p_{\text{ret}}$. This distribution is then fused with the linear classifier distribution $p_{\text{probe}}$ in log space, as defined in Eq.~\eqref{eq:rac}. Where $\alpha$ weights the contribution of the retrieval branch. In the linear probe configurations, the $\alpha$ value was selected on an internal validation split derived from the training datasets, searching over the grid  $\{0, 0.25, 0.5, 0.75, 1\}$.

\begin{equation}
\label{eq:rac}
\text{logits}_{\text{final}} = \log p_{\text{probe}} + \alpha \log p_{\text{ret}}
\end{equation}

In contrast, for the LoRA configurations, $\alpha$ was fixed at $0.5$ due to computational budget. The held-out datasets were never used for tuning. The predicted class corresponds to the highest resulting logit, and the $k$ value was set to 20 in all experiments.

This RAC module is implemented as an exact search in PyTorch. Since embeddings are L2-normalized, cosine similarity reduces to a dot product between the query and the embedding bank, and the top-$k$ neighbors are obtained directly. No approximate nearest neighbor index or external retrieval library such as FAISS (Facebook AI Similarity Search) was used.

\subsection{Held-out Evaluation Protocol and Metrics}
\label{sec:heldout}
To assess model performance under domain shift, each stage is evaluated on a held-out dataset excluded entirely from training, rather than on a test split derived from domains already seen by the model. Each stage uses its own held-out dataset, since the two stages are trained and evaluated as independent classifiers.

Stage 1 is evaluated on the ALL-IDB2 dataset, which contains both leukemic and healthy cells in a balanced proportion (130 and 130 single-cell images). This dataset is not part of the reported pretraining corpus of DinoBloom, which strengthens the evaluation, since it represents a domain unseen by the domain-specific encoder. Stage 1 training used the four remaining datasets, consisting of 122,167 single-cell images. Stage 2 is evaluated on the combination of ALL-IDB2 (representing the ALL class) and AML-Cytomorphology (representing the AML class), because, to the best of the authors' knowledge, no single public dataset contains both subtypes. This results in a test set of 3,424 images (130 ALL and 3,294 AML). Stage 2 training was performed on the remaining two datasets, consisting of 69,400 leukemic single-cell images. In both stages, the held-out datasets are excluded from training to avoid contamination or data leakage. Neither classifier is reused across stages, avoiding data leakage between them. 

As a control experiment, Stage 2 is additionally evaluated under a conventional random 80/20 split stratified by class, using the same encoders and linear classifier. This within-domain split quantifies how much of the reported performance depends on dataset-specific artifacts rather than cell morphology patterns.

Performance is reported as accuracy, macro-averaged recall, and macro-averaged F1-score, defined in Eq.~\eqref{eq:acc}, \eqref{eq:recall}, and \eqref{eq:f1} respectively. Macro averaging assigns equal weight to both classes regardless of their frequency. This is necessary in Stage 2 because the held-out test set is dominated by the AML class, with 3,294 samples against 130 of the ALL class. Without macro averaging, a majority-class predictor would still reach a high recall, hiding the failure on the minority class.

\begin{equation}
\label{eq:acc}
\text{Accuracy} = \frac{TP + TN}{TP + TN + FP + FN}
\end{equation}

\begin{equation}
\label{eq:recall}
\text{Recall}_{\text{macro}} = \frac{1}{C}\sum_{c=1}^{C} \frac{TP_c}{TP_c + FN_c}
\end{equation}

\begin{equation}
\label{eq:f1}
\text{F1}_{\text{macro}} = \frac{1}{C}\sum_{c=1}^{C} 2 \cdot \frac{P_c \cdot R_c}{P_c + R_c}
\end{equation}

Where $TP$, $TN$, $FP$, and $FN$ denote the true positives, true negatives, false positives, and false negatives over the entire evaluation set, $C$ is the number of classes, $TP_c$ and $FN_c$ are the true positives and false negatives of class $c$, and $P_c$ and $R_c$ are the precision and recall of class $c$.

\subsection{Experimental Environment}
All of the results were generated using the same experimental setup: a laptop with Windows 11, an Intel Core i9-13900HX Central Processing Unit (CPU), an RTX 4090 Graphics Processing Unit (GPU), and 32 GB of Random Access Memory (RAM). Embedding extraction, LoRA adaptation, and retrieval search were executed on the GPU, whereas linear probing was executed on the CPU over precomputed embeddings.

Encoder weights were obtained from their official public releases. DinoBloom-B was loaded from the MarrLab repository on the Hugging Face Hub, CLIP from the OpenAI clip-vit-base-patch32 checkpoint through the transformers library, and BiomedCLIP through open\_clip. Embeddings were extracted once per encoder over the full curated corpus and saved to disk, so that linear probing and retrieval augmentation operate on precomputed embeddings. LoRA adaptation requires the raw images, since the encoder is modified during training, and is therefore the only configuration that recomputes embeddings. Embedding extraction over the full corpus required between 6 and 20 minutes per encoder, depending on patch size, while LoRA adaptation required between 4 and 16 minutes per epoch, depending on the encoder and stage.

All the described experiments were executed in Python 3.12.10, using the following library versions: PyTorch 2.5.1 (CUDA 12.1), torchvision 0.20.1, transformers 5.9.0, huggingface\_hub 1.10.1, open\_clip 3.3.0, PEFT 0.20.0, scikit-learn 1.8.0, NumPy 2.4.6, pandas 3.0.3, Pillow 12.2.0, and OpenCV 4.13.0.

\section{Experimental Results}
\label{sec::experimentalresults}
This section presents the experimental results, starting with Stage 1. This benchmark evaluated three encoders under four configurations on a held-out dataset (D5) consisting of 260 single-cell images of leukemic and healthy cells. This evaluation protocol assesses whether Stage 1 can correctly differentiate a leukemic cell from a healthy one. Although the leukemic cells in this held-out set are all of the ALL subtype, these count as leukemic cells at this stage, since the purpose of Stage 1 is to separate leukemic from healthy cells so they can be subtyped in Stage 2. 

\begin{table}[H]
\centering
\caption{Stage 1 performance comparison for leukemic versus non-leukemic cell classification. Evaluated on the held-out dataset (ALL-IDB2, $n=260$).}
\label{tab:stage1_results}
\begin{tabular}{llcccc}
\toprule
\textbf{Model} & \textbf{Metric} & \textbf{Linear probe} & \textbf{Linear probe + RAC} & \textbf{LoRA} & \textbf{LoRA + RAC} \\
\midrule
\multirow{3}{*}{DinoBloom} & Accuracy & 0.8923 & 0.9423 & 0.9308 & 0.9385 \\
 & Recall   & 0.8923 & 0.9423 & 0.9308 & 0.9385 \\
 & F1-score & 0.8923 & 0.9423 & 0.9307 & 0.9384 \\
\midrule
\multirow{3}{*}{BiomedCLIP} & Accuracy & 0.6962 & 0.6923 & 0.7962 & 0.8115 \\
 & Recall   & 0.6962 & 0.6923 & 0.7962 & 0.8115 \\
 & F1-score & 0.6918 & 0.6697 & 0.7886 & 0.8057 \\
\midrule
\multirow{3}{*}{CLIP} & Accuracy & 0.6346 & 0.5577 & 0.9115 & 0.8923 \\
 & Recall   & 0.6346 & 0.5577 & 0.9115 & 0.8923 \\
 & F1-score & 0.6345 & 0.5464 & 0.9112 & 0.8917 \\
\bottomrule
\end{tabular}
\end{table}

As shown in Table \ref{tab:stage1_results}, the best result was achieved by DinoBloom with RAC (0.9423 accuracy), with results ranging from 0.54 to 0.94 depending on the encoder and configuration. The held-out set is uncontaminated for the domain-specific encoder, since D5 is not part of the reported pretraining corpus of DinoBloom, which includes D2, D3, and D4 but not D1. Since D5 and D1 are the only available sources of ALL single-cell images, D1 was assigned to training and D5 to evaluation, as D1 is considerably larger than D5.

\subsection{Impact of Pretraining Specialization}
During Stage 1 evaluation, this study quantified the impact of pretraining specialization. DinoBloom achieved the best overall performance, achieving 0.8923 accuracy under linear probing, which best isolates the encoder's raw ability to classify leukemia or not, since the backbone remains frozen. In contrast, BiomedCLIP achieved 0.6962 accuracy, and CLIP achieved 0.6346 under the same configuration. 

\begin{table}[H]
\centering
\caption{Effect of pretraining specialization on Stage 1, measured with linear probing on the held-out dataset (ALL-IDB2, $n=260$). The encoder is frozen, so differences reflect the pretraining corpus alone. $\Delta$ denotes the gain over the general-purpose baseline (CLIP).}
\label{tab:pretraining_gap}
\begin{tabular}{llccc c}
\toprule
\textbf{Model} & \textbf{Pretraining domain} & \textbf{Accuracy} & \textbf{Recall} & \textbf{F1-score} & \textbf{$\Delta$ Accuracy} \\
\midrule
DinoBloom  & Single-cell hematology      & 0.8923 & 0.8923 & 0.8923 & $+0.2577$ \\
BiomedCLIP & Biomedical image--text      & 0.6962 & 0.6962 & 0.6918 & $+0.0616$ \\
CLIP       & Natural images (general)    & 0.6346 & 0.6346 & 0.6345 & --- \\
\bottomrule
\end{tabular}
\end{table}

As shown in Table~\ref{tab:pretraining_gap}, DinoBloom, which was pretrained on single-cell images, obtained a substantial amount of accuracy gain against the general-purpose encoder CLIP (0.2577). This reinforces the importance of pretraining domain for specialized classification tasks such as leukemia classification using cell-level images. Additionally, BiomedCLIP obtained only a 0.0616 gain over CLIP. Since BiomedCLIP was pretrained on biomedical data, it is expected to outperform CLIP, but the magnitude of this accuracy gain is small, which suggests that general biomedical pretraining is not sufficient when the target task requires identifying specific cytomorphological features. 

\subsection{Can Cost-effective Adaptation Replace Expensive Specialized Pretraining?}
The main question this paper aims to answer is how viable cost-effective techniques such as LoRA or RAC can be for leukemia classification, and whether these techniques can replace the need for specialized pretraining. 

Adding RAC improved performance by 0.0500 for DinoBloom but decreased it for CLIP and BiomedCLIP, suggesting that retrieval can add noise rather than help when the encoder is ineffective at the classification task at hand. On the other hand, the most meaningful finding of these experiments is that LoRA increased CLIP performance by 0.2769, reaching 0.9115 accuracy and surpassing DinoBloom performance with a frozen encoder (0.8923), which directly addresses the research question of this work. These results are presented in Table \ref{tab:lora_gain}.

\begin{table}[H]
\centering
\caption{Effect of LoRA adaptation on Stage 1 accuracy (held-out dataset, $n=260$). Accuracy gain denotes the difference between LoRA and linear probing for each encoder. Retrieval augmentation is not applied in either configuration.}
\label{tab:lora_gain}
\begin{tabular}{lccc}
\toprule
\textbf{Model} & \textbf{Linear probe} & \textbf{LoRA} & \textbf{Accuracy gain} \\
\midrule
DinoBloom  & 0.8923 & 0.9308 & $+0.0385$ \\
BiomedCLIP & 0.6962 & 0.7962 & $+0.1000$ \\
CLIP       & 0.6346 & 0.9115 & $+0.2769$ \\
\midrule
\textbf{Gap (DinoBloom $-$ CLIP)} & \textbf{0.2577} & \textbf{0.0193} & \\
\bottomrule
\end{tabular}
\end{table}

In this specific task, the need for expensive pretraining on hematology images was surpassed by a general-purpose vision model using LoRA, an efficient fine-tuning method. However, DinoBloom with LoRA achieved better performance, indicating that LoRA benefits not only general-purpose encoders but also those with specialized pretraining. Although a specialized encoder can be matched or surpassed using LoRA adaptation, the best overall performance in Stage 1 was achieved by DinoBloom with RAC (0.9423), representing a specialized encoder with a cost-effective method. This suggests that cost-effective techniques remain valuable even when the encoder is already domain-specialized.  

\subsection{Retrieval Augmentation}
The RAC module retrieves the $k$ most similar training embeddings by cosine similarity and aggregates their labels into a similarity-weighted vote. Its effect on Stage 1 was not uniform across encoders. The RAC module improved DinoBloom performance in both linear probe and LoRA settings. In contrast, performance decreased for BiomedCLIP and CLIP under linear probing. After LoRA adaptation, the retrieval branch either improved performance or reduced the magnitude of the degradation. 

\begin{table}[H]
\centering
\caption{Effect of retrieval augmentation on Stage 1 accuracy (held-out dataset, $n=260$). Values denote the change in accuracy when RAC is added to each configuration.}
\label{tab:rac_effect}
\begin{tabular}{lcc}
\toprule
\textbf{Model} & \textbf{Linear probe $\rightarrow$ + RAC} & \textbf{LoRA $\rightarrow$ + RAC} \\
\midrule
DinoBloom  & $+0.0500$ & $+0.0077$ \\
BiomedCLIP & $-0.0039$ & $+0.0153$ \\
CLIP       & $-0.0769$ & $-0.0192$ \\
\bottomrule
\end{tabular}
\end{table}
As shown in Table \ref{tab:rac_effect}, general-purpose encoders did not benefit from the RAC module, which is explained by how the module works. Its usefulness depends on how the encoder organizes the embedding space. When cells are grouped by cytomorphological features, the retrieved neighbors belong to the correct class and the RAC module reinforces the classifier, boosting its performance. When cells are instead grouped by dataset-specific artifacts, such as illumination, staining, or scanner characteristics, the retrieved neighbors come from visually similar sources rather than morphologically similar cells, and the module introduces noise that decreases encoder performance. Therefore, the embedding space needs to encode cytomorphological patterns for the RAC module to be useful. This is consistent with the Stage 1 results, where the domain-specific encoder benefits from retrieval while the general-purpose encoder is degraded by it. 

DinoBloom benefited from the RAC module (+0.0500), whereas CLIP, pretrained on natural images, was degraded by it (-0.0769). BiomedCLIP falls between these two, with a slight performance decrease (-0.0039). RAC behavior after LoRA adaptation supports this interpretation. Once the encoder is adapted to the task, the degradation on CLIP is reduced from -0.0769 to -0.0192, and BiomedCLIP shifts from a decrease to an increase in performance (+0.0153), reinforcing the importance of encoder quality for the RAC module to be effective. 

One methodological difference should be noted when comparing these two settings. The fusion weight $\alpha$ was selected from an internal validation split for the linear probe configurations. In contrast, for the LoRA configurations, the $\alpha$ value was fixed at $0.5$ due to computational budget. Since the retrieval branch carries less weight in the LoRA setting, part of the reduced degradation on CLIP and the shift observed on BiomedCLIP may reflect this inconsistency.   

\subsection{Why Subtype Classification Fails Under Domain Shift}
\label{subsec:5.4}
\begin{figure}[H]
\centering
\includegraphics[width=0.45\textwidth]{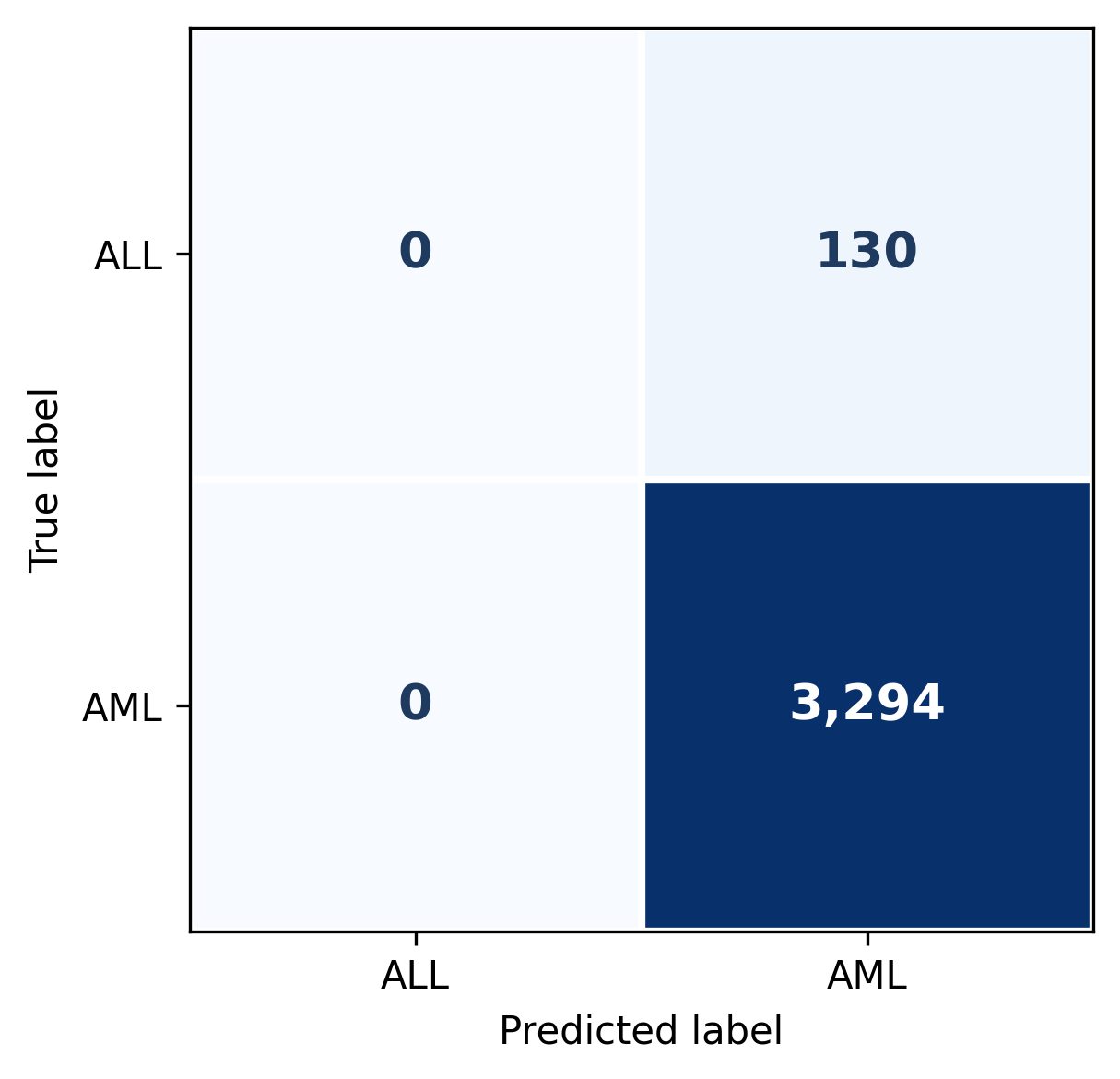}
\caption{Confusion matrix for Stage 2 subtype classification on the held-out datasets, shown for DinoBloom with linear probing. All 130 ALL cells are assigned to the AML class. This pattern is representative of nearly all encoder and adaptation configurations.}
\label{fig:cm_stage2}
\end{figure}

During the Stage 2 experiments, predictions collapsed almost entirely to the AML class. Nearly all configurations classified none of the 130 ALL cells correctly, as shown by the macro-averaged recall values in Table \ref{tab:stage2_results} and by the confusion matrix in Figure \ref{fig:cm_stage2}. This happened even though the training loss reached 0.0000, meaning the model learned the training data perfectly but collapsed on the unseen domains from the held-out datasets. This section examines why this happens.  

\begin{table}[H]
\centering
\caption{Stage 2 performance comparison for ALL versus AML subtype classification. Evaluated on the held-out datasets (ALL-IDB2 and AML-Cytomorphology, $n=3424$). Accuracy reflects the majority class; macro-averaged recall near 0.50 indicates collapse to the AML class.}
\label{tab:stage2_results}
\begin{tabular}{llcccc}
\toprule
\textbf{Model} & \textbf{Metric} & \textbf{Linear probe} & \textbf{Linear probe + RAC} & \textbf{LoRA} & \textbf{LoRA + RAC} \\
\midrule
\multirow{3}{*}{DinoBloom} & Accuracy & 0.9620 & 0.9620 & 0.9617 & 0.9620 \\
 & Recall   & 0.5000 & 0.5000 & 0.4998 & 0.5000 \\
 & F1-score & 0.4903 & 0.4903 & 0.4902 & 0.4903 \\
\midrule
\multirow{3}{*}{BiomedCLIP} & Accuracy & 0.9620 & 0.9620 & 0.9620 & 0.9620 \\
 & Recall   & 0.5000 & 0.5000 & 0.5000 & 0.5000 \\
 & F1-score & 0.4903 & 0.4903 & 0.4903 & 0.4903 \\
\midrule
\multirow{3}{*}{CLIP} & Accuracy & 0.9118 & 0.9118 & 0.9620 & 0.9620 \\
 & Recall   & 0.4850 & 0.4850 & 0.5000 & 0.5000 \\
 & F1-score & 0.4867 & 0.4867 & 0.4903 & 0.4903 \\
\bottomrule
\end{tabular}
\end{table}

There is a specific reason why ALL classifications were biased towards the AML class. Training was limited to one source per subtype: one dataset provided AML samples, and another provided ALL samples, with a synthetic image background. Consequently, the encoder learned the smear background and dataset-specific artifacts rather than morphological cellular features, which are more difficult for a vision encoder to extract. Learning cytomorphological patterns was not necessary, since the models reached a training loss of 0.000 by the second epoch, meaning the optimization did not provide feedback to move beyond dataset-specific shortcuts. This behavior is not explained by class imbalance either. Class weighting was applied in both linear probing and LoRA training, and the loss still converged to zero, suggesting that the model had no problem classifying the minority class during training.

Building on this, we performed an additional experiment on the DinoBloom encoder to assess the RAC module. This revealed that when querying an ALL image in the embedding space, 96.5\% of the retrieved neighbors were AML and only 3.5\% were ALL. This indicates that the encoder was placing the ALL cells on the AML side of the embedding space, before any classifier was applied. A possible explanation is that the held-out ALL dataset lacks a synthetic background, unlike the ALL dataset used for training, so the encoder associates it more closely with the AML images, which have a similar smear background. This behavior is also not attributable to the calibration of the retrieval branch. Increasing the fusion weight up to $\alpha=8$, where retrieval dominates the classifier output, never raises ALL recall above 0.0385 for any of the three encoders. Both findings reinforce that the encoder was learning shortcuts on the background of the cell image rather than cytomorphological patterns. 

To quantify how much of this performance depends on dataset-specific artifacts rather than cell morphology, an additional control experiment was performed. The same models were evaluated under a random 80/20 split within the domain instead of the held-out dataset protocol. To make this dependency explicit, models were evaluated using a linear probe with a frozen encoder. The results presented in Table \ref{tab:stage2_control} show that ALL class recall increased to at least 0.99 in every model. This reinforces the idea that the encoder was learning dataset-specific backgrounds, since the same models collapse to near-zero ALL recall when they are evaluated through a strict held-out dataset evaluation protocol. This is most clearly demonstrated by the CLIP results. A general-purpose model achieved 1.0000 accuracy on a specialized single-cell classification task without any fine-tuning, which contradicts the results in Stage 1 shown in Table \ref{tab:stage1_results}, where CLIP with a frozen encoder achieved only 0.6346 accuracy. Reaching perfect accuracy suggests that CLIP learned the acquisition source rather than hematology features, which is something a non-specialized encoder can do. This is what the evaluation protocol on a held-out dataset uncovered: high performance is achieved under a random split, but when exposing the same model to an unseen domain, performance drops to near-zero ALL recall. 

\begin{table}[H]
\centering
\caption{Stage 2 control experiment. Linear probing is evaluated under a random 80/20 split (within-domain) and under the held-out dataset protocol (cross-domain). ALL recall denotes the proportion of correctly identified ALL cells.}

\label{tab:stage2_control}
\begin{tabular}{lcccc}
\toprule
\textbf{Model} & \multicolumn{2}{c}{\textbf{Random split}} & \multicolumn{2}{c}{\textbf{Held-out dataset}} \\
\cmidrule(lr){2-3} \cmidrule(lr){4-5}
 & Accuracy & ALL recall & Accuracy & ALL recall \\
\midrule
DinoBloom  & 0.9999 & 0.9988 & 0.9620 & 0.0000 \\
BiomedCLIP & 0.9997 & 0.9971 & 0.9620 & 0.0000 \\
CLIP       & 1.0000 & 1.0000 & 0.9118 & 0.0231 \\
\bottomrule
\end{tabular}
\end{table}

This is directly related to what this work means by robustness. Much of the previous work trains and evaluates on the same internal split, but such models' performance can degrade severely when exposed to a new domain, as in a real-world scenario. A model can be trained on data from one hospital and collapse when applied to another, which is why a robust evaluation protocol is needed. Building on this, subtype classification under domain shift could be addressed in two ways. The first would require a dataset containing both ALL and AML single-cell images acquired from the same imaging platform, which would then be used as the held-out evaluation. The second would involve training on additional sources of ALL and AML data, so that the encoder is forced to learn cytomorphological features instead of dataset-specific shortcuts. Neither approach is available with the current public single-cell image datasets, and both could be considered for future work.

\subsection{Comparison with Prior Work}
A direct numerical comparison is not possible in this study, since the literature review found no works using a two-stage cascaded classification pipeline, which would make any comparison unfair. Additionally, some of the previous works perform WBC classification, or address only ALL or AML classification, or evaluate on a test split from a domain already seen by the encoder during training, as discussed in section \ref{subsec:5.4}, where such models are prone to collapse when exposed to a different domain. 

A close study of this work is the benchmark proposed by Ben Rabah and Serag \cite{benrabah2026benchmark}, which quantifies the impact of pretraining specialization across different foundation models. Their evaluation protocol was training on a source dataset and evaluating on a held-out dataset for breast and prostate cancer. However, leukemia was excluded from this cross-dataset evaluation protocol because of inconsistent class definitions across sources, and was instead trained and evaluated using a single dataset and an inner split within domain. The main consequence of using an inner split within domain, as presented in the results of this work, is that the encoder is not evaluated under domain shift, and the encoder's performance may degrade when tested under an unseen domain. This work addresses the obstacle of label harmonization that motivated their study to exclude leukemia classification from the held-out dataset evaluation protocol, enabling cross-dataset training and held-out dataset evaluation for leukemia. However, this made a second obstacle apparent: even after harmonizing labels across heterogeneous datasets, subtype classification remains confounded by dataset-specific artifacts, since no public single-cell dataset provides both subtypes under the same imaging protocol. These two obstacles are layered, and this work documents the second one. 

Wang et al. \cite{wang2026alsnet} report a rigorous evaluation protocol, proposing ALSNet, trained on 180,928 single-cell images from three different imaging platforms and reporting strong performance under held-out dataset evaluation. A relevant difference is the scope; their pipeline performs cell-level morphological classification into nineteen categories, aggregated into case-level predictions from bone marrow smears. In contrast, the present work classifies each cell directly into diagnostic categories such as AML or ALL. Additionally, this study evaluates encoder adaptation through LoRA and retrieval augmentation, which their study does not address.

To the authors' knowledge, no prior study focuses solely on evaluating leukemia detection and subtyping under domain shift and trained across heterogeneous single-cell sources, while additionally implementing encoder adaptation techniques such as LoRA and retrieval augmentation within the same two-stage cascaded pipeline. Therefore, a direct numerical comparison would not be meaningful.

\subsection{Limitations}
Some limitations should be considered when interpreting these results and are described as follows:

\begin{itemize}
    \item \textbf{Size of the Stage 1 held-out set.} The held-out evaluation for Stage 1 contains 260 images. ALL-IDB2 was selected because it is the only dataset containing both classes that lies outside the pretraining corpus of the domain-specific encoder, but its limited size means that small differences in accuracy correspond to a small number of individual cells.

    \item \textbf{Pretraining contamination for the specialized encoder.} Three of the five datasets used in this work are reported within the pretraining corpus of DinoBloom. Both available AML sources fall in this category, so the AML component of the evaluation does not represent a fully unseen domain for that encoder.
\end{itemize}

\section{Conclusions and Future Work}
\label{sec::conclusions}
This work presents a framework for robust leukemia classification using single-cell images across heterogeneous datasets, addressing domain shift by using a synthetic background, label harmonization across five datasets with different class definitions, and a uniform inclusion criterion on non-ambiguous nucleated leukocytes. In Stage 1, under held-out dataset evaluation, pretraining specialization accounts for a 0.2577 accuracy gap between DinoBloom (domain-specific) and CLIP (general-purpose) when both are evaluated with a frozen encoder (linear probing). This gap narrows to 0.0193 once LoRA adaptation is applied. Additionally, CLIP adapted with LoRA (0.9115) surpasses DinoBloom without adaptation (0.8923). The best overall result (0.9423) was achieved by combining the domain-specific encoder with retrieval augmentation and linear probing, without any fine-tuning. Cost-effective adaptation methods can compensate for most of the advantage provided by expensive domain-specific pretraining in this leukemia and subtyping classification task, although the highest performance was still achieved by combining both specialized pretraining with RAC. Therefore, cost-effective adaptation benefits not only general-purpose encoders but also specialized ones, and is a viable alternative if expensive pretraining is not feasible due to time or computational resources. 

The held-out dataset evaluation made visible a limitation that a test split within the domain would not expose in this experimental setting. In Stage 2, subtype classification collapses to the majority class under held-out dataset evaluation, while the same models and data achieve near-perfect accuracy using a random within-domain 80/20 split. This is attributable to each subtype class coming from a single dataset or imaging source during training, which allows the encoder to rely on dataset-specific artifacts instead of cytomorphological features. Therefore, future work should focus on adding more data sources to force the encoder to learn cellular patterns rather than shortcuts based on smear backgrounds. Another direction would be to normalize the backgrounds in all single-cell images across datasets, which could be done by applying the same synthetic background. This would force the encoder to learn cellular patterns instead of dataset-specific artifacts, providing a more consistent comparison.

\acknowledgments 
This work was supported by the Coordinaci\'on Institucional de Investigaci\'on of CETYS Universidad.
\bibliography{report} 
\bibliographystyle{spiebib} 

\end{document}